# Room temperature broadband coherent terahertz emission induced by dynamical photon drag in graphene

J. Maysonnave[1], S. Huppert[1], F. Wang[1], S. Maero[1], C. Berger[2,3], W. de Heer[2], T.B. Norris[4], L.A. De Vaulchier[1], S. Dhillon[1], J. Tignon[1], R. Ferreira[1] and J. Mangeney[1*]

[1] *Laboratoire Pierre Aigrain, Ecole Normale Supérieure, CNRS (UMR 8551), Université P. et M. Curie, Université D. Diderot, 75231 Paris Cedex 05, France*

[2] *School of Physics, Georgia Institute of Technology, Atlanta, Georgia 30332, USA*

[3] *Université Grenoble Alpes / CNRS, Institut Néel, Grenoble, 38042 France*

[4] *Center for Ultrafast Optical Science, University of Michigan, Ann Arbor, Michigan 48109-2099, USA*

**Abstract**

Nonlinear couplings between photons and electrons in new materials give rise to a wealth of interesting nonlinear phenomena [1]. This includes frequency mixing, optical rectification or nonlinear current generation, which are of particular interest for generating radiation in spectral regions that are difficult to access, such as the terahertz gap. Owing to its specific linear dispersion and high electron mobility at room temperature, graphene is particularly attractive for realizing strong nonlinear effects [2]. However, since graphene is a centrosymmetric material, second-order nonlinearities a priori cancel, which imposes to rely on less attractive third-order nonlinearities [3]. It was nevertheless recently demonstrated that *dc*-second-order nonlinear currents [4] as well as ultrafast *ac*-currents [5] can be generated in graphene under optical excitation. The asymmetry is introduced by the excitation at oblique incidence, resulting in the transfer of photon momentum to the electron system, known as the photon drag effect [6]. Here, we show broadband coherent terahertz emission, ranging from about 0.1-4 THz, in epitaxial graphene under femtosecond optical excitation, induced by a dynamical photon drag current. We demonstrate that, in contrast to most optical processes in graphene, the next-nearest-neighbor couplings [7] as well as the distinct electron-hole dynamics are of paramount importance in this effect. Our results indicate that dynamical photon drag effect can provide emission up to 60 THz opening new routes for the generation of ultra-broadband terahertz pulses at room temperature.

Current terahertz (THz) technologies suffer from the lack of compact room temperature THz sources, limiting the proliferation of consumer applications. As a consequence, an important activity in this field is dedicated to developing sources such as photoconductive devices [8,9], quantum cascade lasers [10] or exploring new schemes for THz generation like intracavity difference-frequency generation in mid-infrared quantum cascade lasers [11]. In parallel, important effort is dedicated to the study of new physical properties within novel materials [12,13]. Owing to its gapless electronic band structure, graphene is gaining increasing attention for new developments in the THz domain [14]. In addition, graphene exhibits a large nonlinear optical response arising from the linear carrier energy dispersion, together with the high electron velocity near the Dirac point [2]. Harmonic generation at THz frequencies relying on a third-order nonlinearity has been recently demonstrated in graphene using a femtosecond optical excitation at normal incidence [3]. Although second-order nonlinear effects are generally considerably stronger and therefore of importance in applications, these are forbidden by symmetry as graphene is a centrosymmetric material. Second-order nonlinearities only appear when the photoexcited medium possesses an anisotropy axis, which imposes a preferential direction of motion for the carriers. The photo-excitation can itself, however, introduce an anisotropy direction, related to the in-plane photon momentum $\vec{q}_{//}$. Indeed, second-order nonlinear *dc*-currents have recently been demonstrated in graphene, under monochromatic photo-excitation at oblique incidence and at energy $\hbar\omega < E_F$ [4,15]. These *dc*-currents originate from photon drag or photogalvanic effects [2] and involve only conduction electrons. Recent experiments on graphene under femtosecond pulsed excitation at photon energy $\hbar\omega >> 2E_F$ have shown second harmonic generation in the visible [16] and second-order nonlinear *ac*-currents associated with narrowband THz generation [5] at room temperature.

In this work, we demonstrate and theoretically describe a *dynamical* photon drag effect that permits room temperature broadband, and potentially ultra-broadband, THz emission from femtosecond excited graphene ($\hbar\omega >> 2E_F$). Furthermore we show, by experimental investigations and theoretical modeling, that the transient current results from the intrinsic asymmetry between the conduction and the valence bands (i.e. different energy dispersions and lifetimes). Interestingly, our findings provide a direct probe of the next-nearest-neighbor couplings in graphene that have been difficult to access by experiments [17] and known to induce nontrivial effects in graphene [18]. Our results highlight their essential role in the dynamical photon drag effect.

The investigated multilayer graphene sample is produced by thermal desorption of Si from the C-terminated face of single-crystal 4H-SiC(0001) and contains typically 35-40 layers with non-Bernal rotated graphene planes [19]. It has been shown that each graphene sheet possesses a band structure very similar to that of an individual graphene monolayer [20] and that the first four layers near the substrate are heavily doped, whereas the upper remaining layers are quasi-neutral ($E_F \sim 8$ meV). The experiment, illustrated in Fig. 1c, uses a mode-locked Ti:Sa laser delivering 110 fs optical pulses at a repetition rate of 80 MHz with an optical fluence ranging up to 35 μJ/cm$^2$. It consists of optical pump pulses that excite the graphene sample and optical probe pulses that coherently detect the THz radiation emitted from the graphene sample using conventional electro-optic detection techniques [21]. The central wavelength of the optical pump pulses is 800 nm corresponding to $\hbar\omega = 1.55$ eV $>> 2E_F$. The graphene sample is placed at the focal plane of an aspherical optical lens (effective focal length of 50 mm), so that the incidence angle ϕ of the optical excitation (see Fig. 1c) can be varied by displacing the pump beam position on the surface of the lens. Since electro-optic detection technique is only sensitive to synchronized THz radiation with femtosecond pulses, the incoherent thermal background is suppressed [22]. All measurements are performed at room temperature.

The THz waveform $\vec{E}_{THz}(t)$ generated by exciting the graphene sample with an angle $\phi = 37°$ is reported in Fig. 2a. For this measurement, the pump excitation is s-polarized ($\theta=0$ in Fig 1c) and the electro-optic sampling detection set-up is oriented to detect the projection of $\vec{E}_{THz}(t)$ along the x axis (see Fig 1c). In this configuration, we experimentally verified that, in agreement with theory [23], second-order nonlinear effects in the SiC substrate are cancelled. The observed THz waveform shows a main positive peak with a full-width-at-half maximum of 230 fs, followed by oscillations at longer times. As shown below, these oscillations are induced by the limited bandwidth of the electro-optic detection system [24]. The amplitude spectrum, obtained by the Fourier transform of the temporal electric field waveform, consists of a single broad peak centered at 1.25 THz, as shown in Fig. 2b.

Figure 2e shows that the amplitude of $\vec{E}_{THz}(t)$ scales linearly with the excitation fluence and thus quadratically with the incident optical electric field, indicating a second-order nonlinear process. The electric field peak amplitude reaches 70 mV/cm at an optical fluence of 35 μJ/cm². In order to probe the role of the in-plane photon momentum in the THz generation process, $\vec{E}_{THz}(t)$ is measured for different incidence angles ϕ under s-polarization. In contrast with usual second-order nonlinear processes in two-dimensional systems under s-polarized excitation, that are insensitive to the incidence angle [1], we observe critical changes in the emitted electric field waveform as a function of ϕ. At normal incidence (ϕ=0) no THz signal is detected. For opposite incidence angles, the time-oscillations show reverse polarity, as shown in Fig. 2f and 2g for ϕ=±14°. These results are summarized in Fig 2h that shows the relative field amplitude is proportional to the incident angle and therefore to the in-plane component of the photon momentum. This is a distinctive feature of the dynamical photon drag effect. We show in Fig. 2i that by using a less sensitive but more broadband ZnTe crystal (200 μm-thick), the amplitude spectrum is shifted to higher frequencies with detected components up to 4 THz. This observation indicates that the spectral content of the measured THz radiation is limited by the experimental set-up response.

In order to describe quantitatively the experimental findings and interpret the temporal waveform of the THz signal, we have calculated the time-dependent average current $\langle \vec{j} \rangle$ generated in one graphene layer excited by femtosecond optical pulses, up to the second order in the exciting electric field: $\vec{E}(\vec{r},t) = \text{Re}[E_0 \exp(i\vec{q}.\vec{r} - i\omega t)]\exp(-t^2/2\tau^2)\vec{\varepsilon}$ with $\tau$ the 1/e-width of the laser pulses ($\tau$ = 110 fs). To this end, we calculated the density matrix evolution in the standard perturbation formalism: $i\hbar\, \partial\rho^{(n)}/\partial t = [H_0, \rho^{(n)}] + [V_{dip}, \rho^{(n-1)}] - i\hbar\, \Gamma_n$ up to the second order ($n$=1,2) in the dipolar perturbation $V_{dip} = e\vec{A}.\vec{p}/m_0$. $\Gamma_n$ are phenomenological dampings. In the basis of the graphene eigenstates $|\xi, \lambda, \vec{k}\rangle$ (where $\xi$ is the Dirac-valley index, $\lambda$ the band label and $\vec{k}$ the in-plane wavevector), $\rho^{(0)}$ is purely diagonal and reflects the thermal electron distribution at room temperature before the optical pulse excitation. The THz radiation emitted in the far field is obtained as the time-derivative of the second-order current: $\vec{E}_{THz}(t) \propto \partial \text{Tr}[\vec{p}\rho^{(2)}]/\partial t$. The dominant terms in the calculation of $\vec{E}_{THz}(t)$ involve only the diagonal elements of $\rho^{(2)}$, i.e. the non-equilibrium electron and hole non-linear populations generated by the laser pulse. Fig. 1b schematically shows the anisotropic electron population distribution in the momentum space at the pulse maximum for a s-polarized exciting optical pulse. At normal incidence (Fig 1b, upper panel), currents from electrons with opposite wavevectors cancel each other. At oblique incidence however (Fig. 1b, lower panel), the population distribution is displaced relatively to the centre of the Dirac cone and a net conduction current appears. Nonetheless, the valence current has an opposite direction and compensates exactly the conduction current in the nearest-neighbor (NN) tight-binding approximation, and the total transient photon drag current $\langle \vec{j} \rangle$ vanishes. The inclusion of the next-nearest-neighbor (NNN) coupling in the tight-binding model breaks the mirror symmetry between conduction and valence bands: to the lowest order in **k**, this adds a positive quadratic term to both dispersions, increasing the electron velocity and reducing the hole one. Consistently, the damping coefficients $\Gamma_2^e$ and $\Gamma_2^h$ for electrons and holes are also different in this case. Finally, we obtain that, owing to the electron-hole asymmetry effects, a non-zero photon drag current arises, which is discussed below. The following tight binding parameters have been used: the NN (NNN) energy hopping t=3eV (t'=0.15eV) and an overlap between NN orbitals of s=0.1. The best overall agreement with the experiments was found for $1/\Gamma_2^e$ = 170 fs and $1/\Gamma_2^h$ = 1.025/$\Gamma_2^e$. Once these quantities are fixed, we convolute the calculated $\vec{E}_{THz}(t)$ with the experimental set-up response function, which is a high-frequency filter with a cut-off at ~ 3THz. The complete model becomes thus predictive and allows a quantitative analysis of the various dependencies of the measured THz emission: the incidence angle ϕ, the optical polarization θ and the THz polarization in the x or y directions.

We show in Fig. 2c the calculated transient signal $\vec{E}_{THz} // \vec{x}$ for ϕ=25° and θ=0°. It is in very good agreement with the measured waveform (Fig. 2a) and, importantly, its amplitude is consistent with the experimental value. Moreover, the dynamical photon-drag model nicely reproduces the experimental features in Fig. 2e to 2h: the signature of a second order effect with its dependence on input power and the linear variation of $\vec{E}_{THz}$ with $\vec{q}_{//}$. The broadband nature of this dynamical photon drag effect is highlighted in Fig. 2i, which reports the calculated amplitude spectra of the THz electric field emitted by graphene without any convolution with experimental set-up response: spectral components up to 9 THz are emitted in the far-field region. More, Figure 2j shows the calculated amplitude spectra for an ultra-short optical excitation of 15 fs duration, providing ultra-broadband coherent emission up to 60 THz.

The model can be further tested probing the dependencies with the polarizations of the incoming optical and outgoing THz radiations as reported in Fig. 3b and 3d. When θ is varied, the anisotropic population distributions $\rho^{(2)}(\vec{k},t)$ rotate correspondingly in the **k**-plane, strongly changing the amplitude and direction of the photon-drag current, and therefore of the emitted field $\vec{E}_{THz}(t)$. Dynamical photon drag in graphene has a strong signature for two symmetric polarizations relative to *p* or *s* directions (e.g. θ=45° and θ=135°): the temporal profiles are identical for THz electric fields along the *x* direction (Fig. 3a) and opposite for *y* direction (Fig. 3c). We experimentally confirm this behavior in Figs 3a and 3c that report the measured THz signals along the *x* and the *y* directions respectively at the specific polarization angles of θ=45° and θ=135°. Such an agreement also appears when we compare the Fourier transform $\vec{E}_{THz}(\nu_{THz})$ of the detected THz signal, as shown in the Figs. 2b and 2d and Figs. 3e and 3f. Note that under p-polarized optical excitation, second-order nonlinear effect in SiC substrate contributes to the THz emission for only 30% of its amplitude and at low frequency (<2 THz). We stress that in Figs. 3e and 3f the peak frequency of the THz signal emitted along the *y* direction is higher for *p* than for *s* excitation configuration. Strikingly, the analysis of this frequency shift in the framework of our model permits to extract useful information of the carrier's relaxation processes. Indeed, the computed signal exhibits such a frequency shift *only* if the hole scattering time is taken smaller than the electron one. For the moderate optical fluence used in this study (< 40 μJ/cm$^2$), the involved relaxation rates $\Gamma_2^e$ and $\Gamma_2^h$ are mainly related to carrier-phonon scattering, which efficiently randomizes the direction of carrier wavevector [25,26] making the current vanish.

Our experimental and theoretical study pinpoints the essential physical aspects underlying the dynamical photon-drag effect in graphene excited by femtosecond optical pulses with a photon energy much higher than the Fermi level energy. Moreover, it offers a unique probe of physical properties of graphene such as the next-nearest-neighbor coupling and the distinct dynamics of non-thermal electron and hole population that are otherwise difficult to evaluate. The consideration of the next-nearest-neighbor coupling in the theoretical predictions is in contrast to most optical processes in graphene for which this coupling provides only small corrections to the modeling. Here, this coupling, as well as the asymmetry between the electrons and holes dynamics is intrinsically related to the observed phenomena. Moreover, the new insights provided by our study in the dynamics of the non-equilibrium electron and hole populations during the first hundred of femtoseconds after interband excitation is particularly fascinating since optical gain and population inversion in graphene is possible only in this time window [27]. Furthermore, our work has important implications for THz technology since these results demonstrate that dynamical photon drag effect in multilayer graphene provides an original scheme for coherent pulsed THz emission at room temperature. Our results pave the way to exploit dynamical photon drag effects in many other materials such as graphene-like materials or carbon-based materials.

**Figure Captions:**

**Figure 1 | Dynamical photon drag current induced by non-equilibrium carrier population and experimental schematic. a,** Transient non-thermal electron and hole distributions in the interband regime for an oblique illumination; the electron and hole population distributions are not symmetric with the respect to the center of the Dirac cone. **b,** Non-equilibrium electron population distribution generated by p-polarized femtosecond optical pulses at normal incidence (upper panel) and at oblique incidence (lower panel). The photon momentum was artificially enhanced by a factor 100 in this simulation to make clear the displacement relative to the center of the Dirac cone. The white arrows represent the integrated momentum of the two population distribution lobes. Whereas at normal incidence, the two contributions to the current compensate perfectly, this is not the case at oblique incidence, where a net current is generated. This current changes in direction and amplitude with the direction of the exciting optical electric field, as the latter determines the position of the population distribution lobes around the Dirac cone. **c,** A femtosecond optical pump pulse illuminates the multilayer graphene and creates non-equilibrium electron and hole populations. A transient photon drag current is then generated in the plane of the graphene sheets, which emits a THz pulse. The THz pulse is transmitted through the SiC substrate, collected by an off-axis parabolic mirror and detected in the time-domain using electro-optic sampling in a 1 mm-thick ZnTe crystal.

**Figure 2 | Measured and calculated THz electric field emitted by multilayer graphene. a**, Experimental electric field waveform emitted by the multilayer graphene illuminated by s-polarized femtosecond optical pulses at 800 nm central wavelength under an incidence angle ϕ=37° and its associated spectrum (**b**). **c**, Calculated electric field waveform emitted by the multilayer graphene excited in similar conditions as in (**a**) and its associated spectrum (**d**). **e**, THz electric field amplitude as a function of the optical fluence incident on the multilayer graphene. The red squares are the experimental data and the error bars show the standard deviation associated to noise fluctuations. The dashed line underlines the linear dependence. **f-g**, Time resolved electric field profiles measured for two opposite angles of incidence (ϕ=±14°) under s-polarized optical excitation. **h**, The peak-to-peak amplitude of emitted THz electric field normalized by *(1-r)*, with *r* the amplitude Fresnel coefficient of graphene, as a function of the incidence angle of the femtosecond optical pulses ϕ. The squares are the experimental data, the error bars show the standard deviation associated to noise fluctuations and the black solid line is the best-fit sinus curve. **i**, Amplitude spectra of the THz electric field emitted by graphene excited by 110 fs optical pulses measured with a 1mm-thick ZnTe crystal (green dashed curve), with a 200 μm-thick ZnTe crystal (green solid curve) and calculated without any convolution with the experimental set-up response (red curve). **j**, Calculated amplitude spectrum of the THz signal emitted by graphene excited by 15 fs optical pulses.

**Figure 3 | Evolution of the THz electric field with the orientation of the linear polarization of the optical pulses.** Measured **(a)** and calculated **(b)** electric field waveforms emitted by the multilayer graphene excited by linearly polarized optical pulses and detected in the x direction for two different orientations of the optical polarization θ=45° (solid line) and θ=135° (dashed line). **(c,d)** Electric field waveforms emitted by multilayer graphene

excited by linearly polarized optical pulses and detected in the y direction for two different orientations of the optical polarization θ=45° (solid line) and θ=135° (dashed line). Spectra of the experimental **(e)** and calculated **(f)** transient electric field emitted by multilayer graphene illuminated by *s*-polarized (green) and *p*-polarized (red) femtosecond optical pulses and detected in the x direction.

**Figure 1**

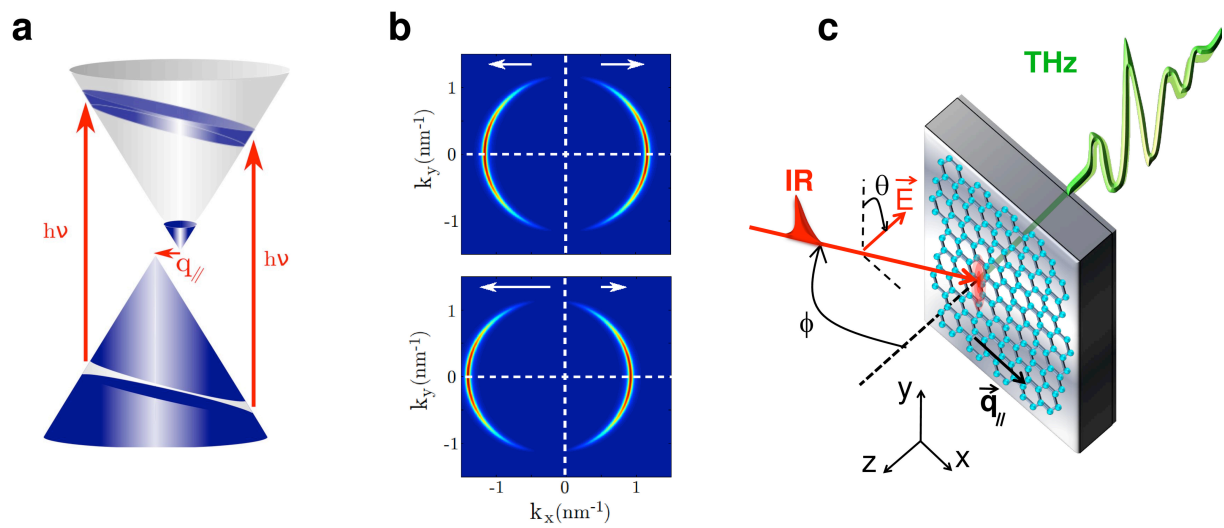

**Figure 2**

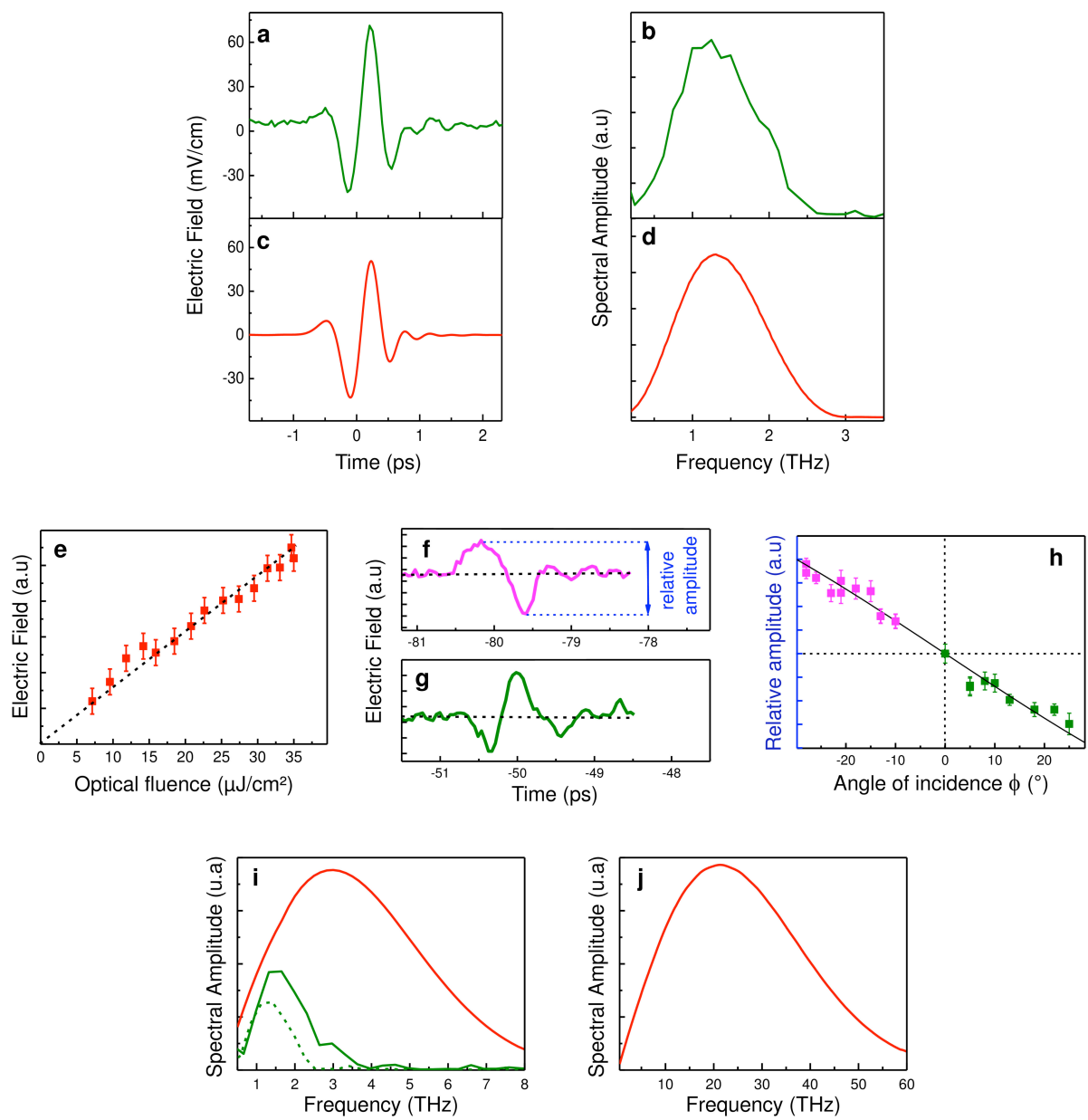

**Figure 3**

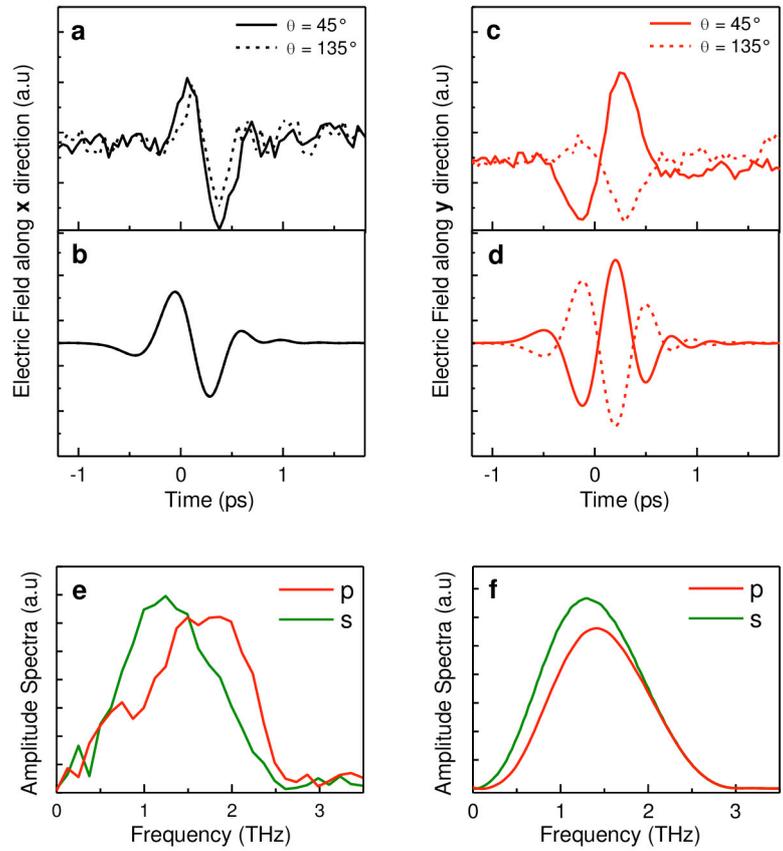